\documentclass[11pt,oneside,english,reqno,a4paper]{amsart}

\usepackage{graphicx}
\usepackage{mathrsfs}
\usepackage[numbers]{natbib}
\usepackage[fit]{truncate}
\usepackage{amssymb,amsmath,amsfonts}
\usepackage[english]{babel}
\usepackage[utf8]{inputenc}
\usepackage[compatible]{algpseudocode}
\usepackage{float}
\usepackage{diagbox}
\usepackage[hidelinks]{hyperref}
\usepackage{hhline}

\newcommand{\R}{\ensuremath\mathbb{R}}
\newcommand{\N}{\ensuremath\mathbb{N}}

\newcommand{\E}{\ensuremath\mathbb{E}}

\newcommand{\el}{\eta_\text{l}}
\newcommand{\es}{\eta_\text{s}}
\newcommand{\els}{\eta_\text{ls}}
\newcommand{\bl}{\beta_\text{l}}
\newcommand{\bs}{\beta_\text{s}}
\newcommand{\bls}{\beta_\text{ls}}
\newcommand{\bel}{\boldsymbol{\eta}_\text{l}}
\newcommand{\bes}{\boldsymbol{\eta}_\text{s}}
\newcommand{\bels}{\boldsymbol{\eta}_\text{ls}}
\newcommand{\by}{\boldsymbol{y}}
\newcommand{\bt}{\boldsymbol{t}}
\newcommand{\bth}{\boldsymbol{\vartheta}}
\newcommand{\bT}{\boldsymbol{T}}
\newcommand{\bd}{\boldsymbol{\delta}}
\newcommand{\Xl}{\boldsymbol{X}_\text{l}}
\newcommand{\Xs}{\boldsymbol{X}_\text{s}}
\newcommand{\Xls}{\boldsymbol{X}_\text{ls}}

\newcommand{\Tstrut}{\rule{0pt}{2.6ex}}
\newcommand{\Bstrut}{\rule[-0.9ex]{0pt}{0pt}} 

\DeclareMathOperator{\Var}{\text{Var}}
\DeclareMathOperator{\argmax}{arg\,max}
\DeclareMathOperator{\argmin}{arg\,min}
\DeclareMathOperator{\mat}{\text{Mat}_\R}

\numberwithin{equation}{section}

\begin{document}

\title[Extending Joint Models for Gradient Boosting]{Extension of the Gradient Boosting Algorithm for Joint Modeling of Longitudinal and Time-to-Event data}
\author{Colin Griesbach}
\author{Andreas Mayr}
\author{Elisabeth Waldmann}
\thanks{The work on this article was supported by the Interdisciplinary Center for Clinical Research (IZKF) of the Friedrich-Alexander-University Erlangen-Nürnberg (Project J61).}

\address{Department of Medical Informatics, Biometry, and Epidemiology, Friedrich-Alexander-University Erlangen-Nürnberg, Waldstr. 6, D-91054 Erlangen}
\email{colin.griesbach@fau.de}
\address{Department of Medical Biometry, Informatics and Epidemiology, Faculty of Medicine, University of Bonn}
\email{mayr@uni-bonn.de}
\address{Department of Medical Informatics, Biometry, and Epidemiology, Friedrich-Alexander-University Erlangen-Nürnberg, Waldstr. 6, D-91054 Erlangen}
\email{elisabeth.waldmann@fau.de}

\maketitle

\begin{abstract}
In various data situations joint models are an efficient tool to analyze relationships between time dependent covariates and event times or to correct for event-dependent dropout occurring in regression analysis. Joint modeling connects a longitudinal and a survival submodel within a single joint likelihood which then can be maximized by standard optimization methods. Main burdens of these conventional methods are that the computational effort increases rapidly in higher dimensions and they do not offer special tools for proper variable selection. Gradient boosting techniques are well known among statisticians for addressing exactly these problems, hence an initial boosting algorithm to fit a basic joint model based on functional gradient descent methods has been proposed. Aim of this work is to extend this algorithm in order to fit a model incorporating baseline covariates affecting solely the survival part of the model. The extended algorithm is evaluated based on low and high dimensional simulation runs as well as a data set on AIDS patients, where the longitudinal submodel models the underlying profile of the CD4 cell count which then gets included alongside several baseline covariates in the survival submodel.
\end{abstract}

\section*{Introduction}\label{sec_intro}
Joint models turned out to be a powerful approach to analyzing data where event times
are measured alongside a longitudinal outcome and were first suggested by Wulfsohn
and Tsiatis \cite{Wulfsohn.1997}. When dealing with such a very common data structure, one naive approach would be separate modeling, i.e. fitting some suitable longitudinal model and in addition a Cox regression for event times. The disadvantages of this approach are that separate modeling neither corrects for event-dependent dropout in longitudinal analysis, nor quantifies the relation between a time-dependent covariate and the risk for an event in survival analysis which has been shown based on simulations by Guo and Carlin \cite{Guo.2004}. To overcome these issues, various modeling approaches have been proposed, e.g. the extended Cox model including time-dependent covariates or two stage approaches, where longitudinal models are fit in order to carry out survival analysis. These approaches however happen to produce biased results like Rizopoulos \cite{Rizopoulos.2012} and Sweeting \cite{Sweeting.2011} showed for the extended Cox model, respectively the two stage approach. The solution is combining both the survival and longitudinal submodel within one single joint likelihood. An introduction for this joint modeling framework can be found in \cite{Rizopoulos.2012} including the \texttt{JM} package discussed in Rizopoulos \cite{Rizopoulos.2010}. Furthermore Tsiatis and Davidian \cite{Tsiatis.2004} give a summary of joint model evolution up to 2004.

One main drawback in current joint modeling estimation approaches is that they are inappropriate for high dimensional data, especially when the number of covariates exceeds the number of observations, and lack clear strategies to address variable selection properties. Thus, Waldmann \textit{et al.} \cite{Waldmann.2017} proposed an estimation procedure based on gradient boosting techniques. Evolved from machine learning as an approach to classification problems initially proposed by Freund and Schapire \cite{Freund.1996}, gradient boosting can be seen as conventional gradient descent techniques transferred into function space. This connection between boosting and functional gradient descent was first discovered by Breiman \cite{Breiman.1998, Breiman.1999}. For a general summary of the statistical perspective on boosting, see Bühlmann and Hothorn \cite{Buehlmann.2007}, for theoretical convergence results using specific loss functions Bühlmann and Yu \cite{Buehlmann.2003}.

Based on the work by Waldmann \textit{et al.} \cite{Waldmann.2017} it is for the first time possible to estimate and select high dimensional joint models with respect to minimizing the prediction error. However, their initial joint model boosting approach did not include baseline covariates exclusively for the survival submodel. Aim of this work is to extend the discussed algorithm by incorporating another boosting step focusing solely on the newly introduced survival predictor, as well as updating and improving the simulation method by a different approach to generate event times.

The remainder of this paper is structured as follows: Section \ref{sec_model} gives a detailed description of the considered joint model, while boosting in general and the extended boosting algorithm for joint models are discussed in Section \ref{sec_boosting}. Sections \ref{sec_simulation} and \ref{sec_data} deal with applying the algorithm to different setups of simulated data as well as to the AIDS dataset included in the \texttt{JM} package (see Abrams \textit{et al.} \cite{Abrams.1994}). Finally the results and possible extensions are discussed.

\section{The Joint Model}\label{sec_model}
In the following we will give a detailed description of the two submodels in order to formulate the joint likelihood function.

Suppose we have $n\in\N$ individuals with $n_i$ longitudinal measurements per individual $i=1,\dots,n$. The longitudinal part has the form
\begin{equation*}
	y_{ij} = \el(x_{\text{l}ij}) + \els(x_{\text{ls}i},t_{ij}) + \varepsilon_{ij},
\end{equation*}
where $y_{ij}$ denotes the $j$th longitudinal outcome of the $i$th individual. The \textit{longitudinal predictor} $\el$ includes linear functions of $p_\text{l}\in\N$ possibly time-varying covariates and an intercept, thus $\el(x) = \beta_0+\beta_\text{l}^Tx$ with $(\beta_0,\beta_\text{l})\in\R^{p_\text{l}+1}$. The \textit{shared predictor} $\els$ contains in addition to linear functions of $p_\text{ls}\in\N$ time-constant covariates also individual specific random effects and a fixed linear time effect. Hence we have $\els(x,t) = \gamma_0+\beta_\text{ls}^Tx+(\beta_t+\gamma_1)t$ with $(\beta_\text{ls},\beta_t)\in\R^{p_\text{ls}+1}$ and $\gamma_0,\gamma_1\in\R$. The error terms $\varepsilon_{ij}$ are assumed to follow a normal distribution with $\E[\varepsilon_{ij}]=0$ and $\Var(\varepsilon_{ij})=\sigma^2>0$.

The survival model is given by the hazard function
\begin{equation*}
	\lambda_i(t|\es,\els,\alpha,\lambda_0) = \lambda_0 \exp(\es(x_{\text{s}i}) + \alpha\els(x_{\text{ls}i},t))
\end{equation*}
with $\es(x)=\bs^Tx, \ \bs\in\R^{p_\text{s}}$ denoting linear functions of $p_\text{s}\in\N$ covariates fixed in time and a baseline hazard $\lambda_0$ chosen to be constant. Furthermore the $\alpha$-scaled shared predictor introduced in the longitudinal model reappears in the survival part. Thus $\alpha$, also called the association parameter, quantifies the relation between the two submodels.

Now we can formulate the log-likelihood. Let $T_i$ denote the event time of individual $i$ with censoring indicator $\delta_i$ and introduce the vectors $\by$ and $(\bT,\bd)$ as the collection of all longitudinal and survival measurements. Set the parameter vector to be $\bth := (\beta_0,\bl,\bs,\bls,\beta_t,\alpha,\lambda_0,\sigma^2)$, so we can calculate the log-likelihood
\begin{align}\label{eq_likelihood}
	\begin{split}
		\ell(\bth|&\by,\bT,\bd)\\
		&= \sum_{i=1}^{n} \Bigg\{ \sum_{j=1}^{n_i} \Bigg(-\log(\sqrt{2\pi\sigma^2}) - \frac1{2\sigma^2}\big(y_{ij} - \eta_\text{l}(x_{\text{l}ij}) - \eta_\text{ls}(x_{\text{ls}i},t_{ij})\big)^2\Bigg) \\
		&\quad + \delta_i\big( \log\lambda_0 + \eta_\text{s}(x_{\text{s}i}) + \alpha\eta_\text{ls}(x_{\text{ls}i},T_i)\big)\\
		&\quad - \lambda_0 \int_0^{T_i} \exp\big(\eta_\text{s}(x_{\text{s}i}) + \alpha\eta_\text{ls}(x_{\text{ls}i}, u)\big)du \Bigg\},
	\end{split}
\end{align}
since the error terms $\varepsilon_{ij}$ are normal distributed.
Observe $\bth\in\R^{p_\text{l}+p_\text{s}+p_\text{ls}+3}\times (0,\infty)^2=:\Theta$ so $\ell$ can be seen as a mapping $\ell\colon\Theta\to\R$. We are interested in the maximum likelihood estimator
\begin{equation}\label{eq_mle_problem}
\hat{\bth} := \underset{\bth\in\Theta}\argmax \ \ell(\bth|\by,\bT,\bd)
\end{equation}
for given longitudinal data $\by$ and event times $(\bT,\bd)$. This maximization is achieved via component-wise gradient boosting methods discussed in the following section.

\section{Methods}\label{sec_boosting}
In this section we give a brief overview on the concept of gradient boosting in general and then focus on developing an approach for the class of joint models with regards to the problem proposed in formula (\ref{eq_mle_problem}).

\subsection{Componentwise Gradient Boosting}
Though having its roots in machine learning as an approach to classification problems \cite{Freund.1996}, from a mathematical point of view, gradient boosting can be seen as conventional steepest descent algorithms transformed into function space \cite{Breiman.1998, Breiman.1999, Mason.1999}. If we want to minimize a function $f\colon\R^n\to\R$ one basic iterative method is to compute the negative gradient -$\nabla f$ evaluated at the current position. We then \textit{walk} the by -$\nabla f$ indicated direction for a specific steplength $\nu$ to receive a new value and eventually converge into a local minimum of the cost function $f$. So we have
\begin{equation}\label{eq_general_gda}
x_\text{new} = x_\text{old} - \nu \nabla f(x_\text{old}),
\end{equation}
where the optimal $\nu$ is usually chosen via line search methods, see e.g. \cite{Wolfe.1969}.

This idea remains the same for gradient boosting. We are now interested in finding an optimal predictor function $\eta$ that minimizes a pre-chosen loss function $\rho$ between the evaluation of the predictor at the specific covariates $\eta(x)$ and the given data $y$. Popular examples for this loss function are the euclidean distance or the negative likelihood. The task is now to find an optimal function $\eta$ which minimizes the functional $\rho(y,\eta(x))$. In addition the predictor $\eta(x)$ is split into several \textit{baselearners} $\eta(x) = h_1(x)+\dots+h_p(x)$. Usually one baselearner models the effect of one single covariate; e.g. baselearners in a regular linear regression model would take the form $h_r(x_r) = \beta_r x_r$ as the linear effect of the $r$th covariate.

For componentwise gradient boosting it is crucial to compute the functional derivative
\begin{equation*}
	u = \frac{\partial\rho}{\partial\eta}(y,\eta(x))
\end{equation*}
in order to fit the single baselearners $h_1,\dots,h_p$ separately to the negative gradient $-u$. This can be seen as an approximation of the gradient in order to obtain a \textit{steepest descent update} in analogy to formula (\ref{eq_general_gda}). Now only the best performing baselearner $h^*$ is updated to receive the new predictor
\begin{equation*}
	\eta_\text{new}(x) = \eta_\text{old}(x) + \nu h^*(x).
\end{equation*}
The steplength $\nu = c \nu^*$ is chosen as the optimal $\nu^*$ to minimize the loss function $\rho$ but shrunken by a constant factor $c<1$ in order to give every baselearner an equal chance to be picked and avoid overfitting. Since the optimal steplength for gradient boosting methods using the euclidean distance in fact equals 1, in the literature a famous choice for the shrunken steplength is $\nu = 0.1$. Gradient boosting techniques are also capable of handling smooth effects, see Schmid and Hothorn \cite{Schmid.2008}. A wide class of models and baselearners is included in the \texttt{R} package \texttt{mboost} available on CRAN, a tutorial to \texttt{mboost} can be found in \cite{Hofner.2014}. 

\subsection{Boosting Joint Models}
We now want to formulate the explicite algorithm for optimizing the likelihood (\ref{eq_likelihood}) and give a detailed description. The algorithm in appendix is the proposed algorithm to carry out gradient boosting for joint modeling. The implementation of the algorithm as well as the simulation function discussed in the next section are provided with the new \texttt{R} add-on package \texttt{JMboostE} which source code is hosted openly on\\ \texttt{http://www.github.com/cgriesbach/JMboostE}.

Since both sub-models contain different, but not disjunct compositions of the predictors $\el,\es$ and $\els$, a seperate boosting step for every sub-model would not work out. Thus the general concept is to create an outer loop, in which every predictor is boosted separately. In a fourth step, the parameters $\alpha, \lambda_0$ and $\sigma^2$ are updated according to the current likelihood. This approach is an extension of boosting methods in multiple dimensions including nuisance parameters discussed in \cite{Mayr.2012}. Since the exact sequence of substeps in the three main iterations is basically the same, we give a general explanation instead of carrying out every single step.

\textbf{Initializing base-learners.} Although the computation of the gradient depends on the exact definition of the base-learner functions $h_{\bullet 1},\dots,h_{\bullet p_\bullet}$, there are no restrictions to their specific form. Yet it is necessary to account for the different degrees of freedom as e.g. non-linear base-learners approximated via splines have a higher chance to be picked in the variable selection process otherwise as discussed in Hofner \textit{et al.} \cite{Hofner.2011}. In addition note, that the random effects are formulated as a single base-learner appearing in the step for the longitudinal predictor and are updated all at once, if selected.

\textbf{Computing the gradient.} Primary goal in every boosting iteration is computing the gradient $u$ as the functional derivative of the loss function with regard to the current predictor. This gradient indicates the \textit{direction} which the base learners need to be fit to in order to obtain the optimal update. Since we have chosen the negative log-likelihood $-\ell$ as our loss function $\rho$, the gradient for the survival part takes the form
\begin{align}\label{eq_us}
	\begin{split}
		\boldsymbol{u}_{\text{s}} &= -\frac{\partial \rho}{\partial \es} = \frac{\partial \ell}{\partial \es}\\
		&=\Bigg( \delta_i - \lambda_0\exp(\es)\frac{\exp(\alpha\els(\cdot,T_i)) - \exp(\alpha\eta_{\text{ls}-ti})}{\alpha(\beta_t + \gamma_1)}\Bigg)_{i=1,\dots,n},
	\end{split}
\end{align}
where $\eta_{\text{ls}-t}(x) = \gamma_0 + \tilde{\beta}_\text{ls}^Tx$ denotes the time-independent part of the shared predictor. A detailed derivation of formula (\ref{eq_us}) can be found in the Appendix. For the shared predictor a straight forward computation leads to the desired gradients as well. An explicit formulation can be found in the algorithm, the Appendix of \cite{Waldmann.2017} shows the calculation. 

\textbf{Stopping iterations.} Main tuning parameter for the algorithm are the stopping iterations $m_\text{l}$, $m_\text{s}$ and $m_\text{ls}$. Since the single predictors are updated separately, the best working triple $(m_\text{l}^*,m_\text{s}^*,m_\text{ls}^*)$ has to be found. This is done via cross validation where the coefficients are fit to a set of training data, while the evaluation of the different choices for the stopping iterations is performed on the remaining individuals. The training and test datasets are received by splitting the data randomly in different subsets.

\textbf{Steplengths.} Boosting three predictor functions not only leads to three different upper boundaries for the number of total iterations, but also to three different choices for the steplength in each updating scheme. For the steplengths used in the longitudinal and shared predictor we use the established steplength $\nu_\text{l} = \nu_\text{ls} = 0.1$ mentioned in the previous section. The steplength of the survival submodel is chosen to be $\nu_\text{s} = 0.3$ since the algorithm needs more iterations to converge into the overall maximum of the survival likelihood. This choice is of course not compulsory as lowering the weakness of the learner technically reduces estimation accuracy. One has to find a reasonable choice for $\nu_\text{s}$ with respect to a feasible computational effort and minimal trade-off between computation time and quality of the results.

\textbf{Time and random effects.} Although the time and random effects are part of the shared predictor $\els$ they are estimated solely on longitudinal data alongside with the predictor $\el$. Since it has been shown in \cite{Hofner.2008} that gradient boosting is not capable of handling time effects in survival analysis, this choice is substantial for the estimates' reliability. They nevertheless have an important impact on the joint model through their $\alpha$-scaled appearance in the hazard function.

\section{Simulation}\label{sec_simulation}
In order to evaluate the method proposed in Section \ref{sec_boosting}, we wrote a simulation algorithm which simulates data in context of the scenario discussed in Section \ref{sec_boosting}. The simulation evolved from the approach described in the Appendix of \cite{Waldmann.2017} but comes with some crucial changes in the part where event-times are generated. This leads to bias reduction, as the representation of the true distribution of event times is more accurate.

\subsection{Data generation}
The algorithm in Appendix \ref{seq_algo_sim} briefly describes the me\-thod to simulate data joint modeling is suitable for. We now want to give a detailed description to the single steps. 

\textbf{Time points}. The setup simulates a survey over $n_i$ years where every individual attends the survey randomly once a year. The first longitudinal measurement time is set $t_{i1}=0$ to avoid \textit{missing not at random} censoring, which would occur by ignoring extremely quick occurring events completely.

\textbf{Covariate matrices}. We have $\Xl \in \mat(n\cdot n_i, p_\text{l})$ and $\Xls \in \mat(n\cdot n_i, p_\text{ls})$ where the first column of $\Xl$ is the unit vector $\boldsymbol{1}$ and the last column of $\Xls$ the time vector $\bt$ with respect to the design of $\bl$ and $\bls$. Note that although containing data for multiple longitudinal time points, $\Xls$ only inherits time independent covariates. For the survival covariates we have $\Xs \in \mat(n, p_\text{s})$

\textbf{Event times}. The event times $(\bT,\bd)$ are generated via inversion sampling. Given the distribution function $F(t) = P(T<t)$ of the event times, we can generate event times with distribution function $F$ by drawing uniformly distributed random numbers $u_i \sim \mathcal{U}([0,1])$ and setting $T_i = F^{-1}(u_i)$. The inverse $F^{-1}$ has a closed form solution and can be obtained via a straight forward computation. We consider an event time as censored, if it exceeds the last longitudinal measurement time, since this indicates the end of the survey and the individual is not longer under observation.

\subsection{Results}

Overall, we considered two different simulation setups. One low dimensional situation mimicking a more common dataset and one high dimensional to evaluate performance in cases where $p>n\cdot n_i$ holds.

In both cases two different data sets were generated, training data with $n=500$ and test data with $n'=1000$ individuals. In the first step a model was fit to the training data, while the test data was used to evaluate this model's performance. Fitting and evaluating the model was done via grid search. Based on a three dimensional grid containing possible triples of stopping iterations, a model was fit to the training data for each of those triples. The parameter estimates were then used to compute the overall likelihood function (\ref{eq_likelihood}) for the test data, in order to evaluate the performance of every predefined triple of stopping iterations.

We used the grids $\{60,90,\dots,300\}^3$ and $\{25,50,\dots,175\}^3$ for the low and high dimensional setup each. In both cases the selected stopping iterations did not exceed the grid, which means they did not concentrate on the upper boundary for any predictor. Tables \ref{table_coef} and \ref{table_als} summarize the estimates of the single coefficients as well as the parameters $\alpha$, $\lambda_0$ and $\sigma^2$ for both simulation runs S1 (low dimensional) and S2 (high dimensional), which are discussed in more detail in the following two subsections. The variable selection properties are depicted in Table \ref{table_fp} where TP/FP stands for true/false positive indicating the rate of correctly/incorrectly picked variables.

\begin{table}[t]
	\centering
	\caption{Coefficient estimates}
	\label{table_coef}
	\setlength{\tabcolsep}{3pt}
	\begin{tabular}{|l|lll|}
		\hline
		& $\beta_0$ & $\beta_\text{l1}$ & $\beta_\text{l2}$ \Tstrut\Bstrut\\
		\hline
		Truth & 2 & 1 & -2 \Tstrut\\
		S1 (sd) & 2.006 (0.012) & 0.996 (0.009) & $-1.997$ (0.008)\\
		S2 (sd) & 2.007 (0.017) & 0.994 (0.010) & $-1.995$ (0.010) \Bstrut\\
		\hhline{|====|}
		& $\beta_\text{s1}$ & $\beta_\text{s2}$ & $\beta_\text{s3}$ \Tstrut\Bstrut\\
		\hline
		Truth & $-1$ & 2 & 1 \Tstrut\\
		S1 (sd) & $-0.916$ (0.090) & 1.822 (0.105) & 0.906 (0.087)\\
		S2 (sd) & $-0.701$ (0.092) & 1.492 (0.084) & 0.693 (0.080) \Bstrut\\
		\hhline{|====|}
		& $\beta_\text{ls1}$ & $\beta_\text{ls2}$ & $\beta_t$ \Tstrut\Bstrut\\
		\hline
		Truth & 1 & $-2$ & 2 \Tstrut\\
		S1 (sd) & 1.000 (0.007) & $-1.999$ (0.007) & 1.980 (0.029)\\
		S2 (sd) & 0.986 (0.016) & $-1.983$ (0.020) & 1.972 (0.045) \Bstrut\\
		\hline
	\end{tabular}
\end{table}

\begin{table}[b]
	\centering
	\caption{Estimates for $\alpha$, $\lambda_0$ and $\sigma^2$}
	\label{table_als}
	\begin{tabular}{|l|lll|}
		\hline
		& $\alpha$ & $\lambda_0$ & $\sigma^2$ \Tstrut\Bstrut\\
		\hline
		Truth & 0.5 & 0.1 & 0.1 \Tstrut\\
		S1 (sd) & 0.469 (0.038) & 0.133 (0.025) & 0.093 (0.003)\\
		S2 (sd) & 0.412 (0.033) & 0.217 (0.030) & 0.090 (0.006) \Bstrut\\
		\hline
	\end{tabular}
\end{table}

\begin{table}[h]
	\centering
	\caption{Variable selection properties}
	\label{table_fp}
	\begin{tabular}{|l|llllll|}
		\hline
		& $\text{TP}_\text{l}$ & $\text{FP}_\text{l}$& $\text{TP}_\text{s}$& $\text{FP}_\text{s}$& $\text{TP}_\text{ls}$ & $\text{FP}_\text{ls}$ \Tstrut\Bstrut\\
		\hline
		S1 & 1.00 & 0.012 & 1.00 & 0.598 & 1.00 & 0.990 \Tstrut\\
		S2 & 1.00 & 0.013 & 1.00 & 0.035 & 1.00 & 0.022 \Bstrut\\
		\hline
	\end{tabular}
\end{table}

\subsubsection{Simulation 1: Low dimensional setting}\label{sssec_low_dimensional}
\begin{figure*}[h]
	\includegraphics[scale=0.8]{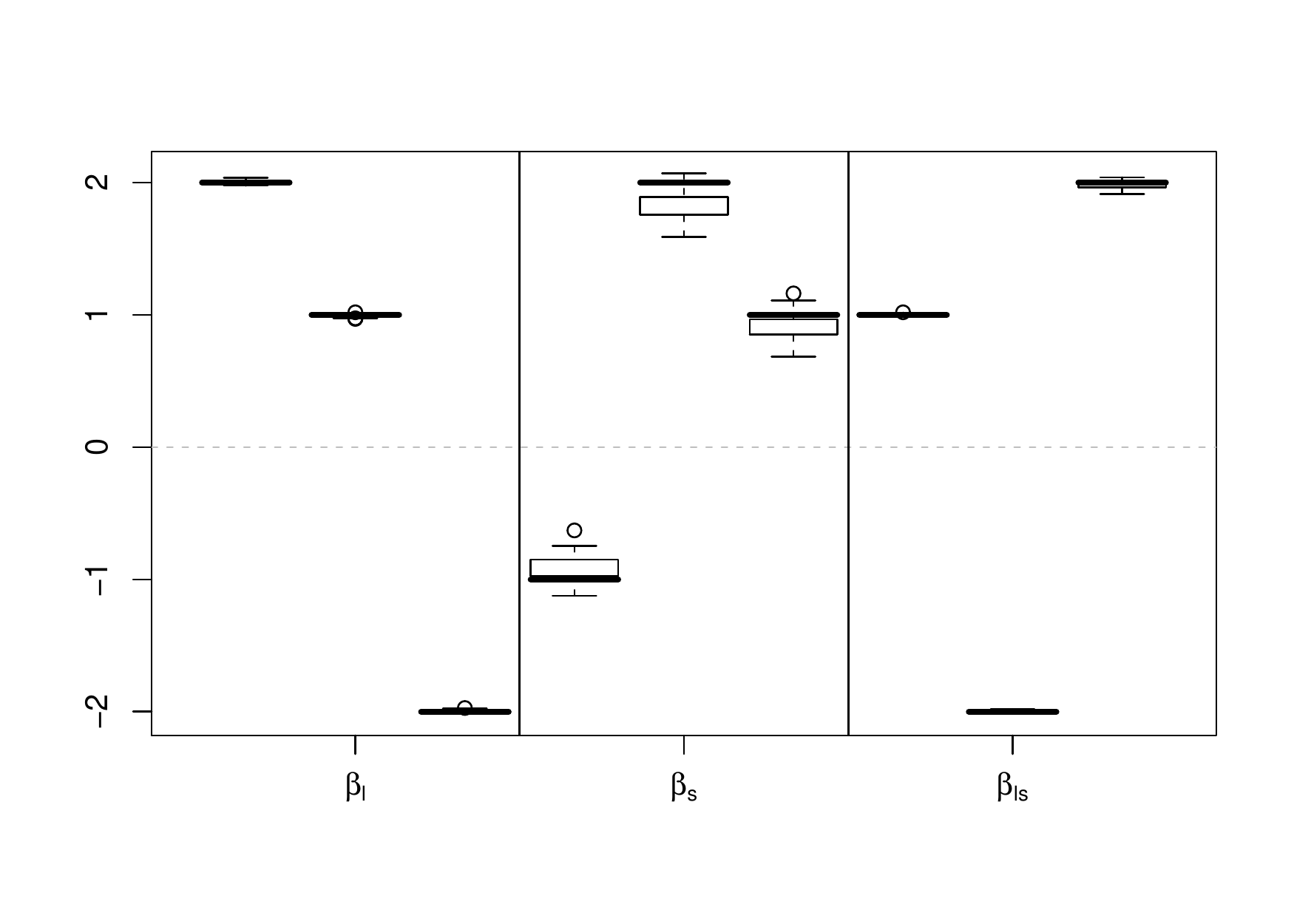}
	\caption{Coefficient estimates for 100 runs in Simulation 1 - thick lines indicate true values}
	\label{fig_s1_coef}
\end{figure*}

For the first simulation $n=500$ individuals with $n_i = 5$ longitudinal measurements per individual were simulated. We chose the coefficients
\begin{equation*}
	\bl :=
	\begin{pmatrix}
		2 \\ 1 \\ -2
	\end{pmatrix}, \quad
	\bs :=
	\begin{pmatrix}
		-1 \\ 2 \\ 1
	\end{pmatrix}, \quad
	\bls :=
	\begin{pmatrix}
		1 \\ -2 \\ 2
	\end{pmatrix},
\end{equation*}
where the first component of $\bl$ is the intercept $\beta_0$ and the last component of $\bls$ the time effect $\beta_t$. The remaining parameters were set as
\begin{equation*}
	\alpha := 0.5, \quad \lambda_0 := 0.1, \quad \sigma^2 := 0.1.
\end{equation*}
Furthermore we added 15 non-informative covariates, six in each predictor, leading to 24 covariates overall. The coefficient estimates of 100 simulation runs in total can be seen in Figure \ref{fig_s1_coef}.
The estimates are very close to the real values, which also holds for the remaining parameters. The coefficients of the survival predictor $\es$ experience shrinkage which leads to a tiny trade-off between $\alpha$ and $\lambda_0$. The mean stopping iterations for each predictor are
\begin{equation*}
	m_\text{l} = 150.6, \quad m_\text{s} = 260.7, \quad m_\text{ls} = 176.1
\end{equation*}
(averaged over all 100 simulation runs). However, in a low dimensional setup false positive rates tend to be higher, in particular non-informative covariates are selected at a rate of $59.8\%$ in the survival and $99.0\%$ in the shared predictor. Because of the design of gradient boosting algorithms, which focuses on predictive risk minimization rather than variable selection, these methods tend to have higher false positive rates in low dimensional settings. Since the random effects are boos\-ted alongside the longitudinal predictor, the algorithm tends to pick these instead of non-informative longitudinal covariates, hence we have a false positives rate of $1.2\%$ in this case. True positives, on the other hand, are recognized 100\% of the time.

\begin{figure}[h]
	\centering
	\includegraphics[scale=0.5]{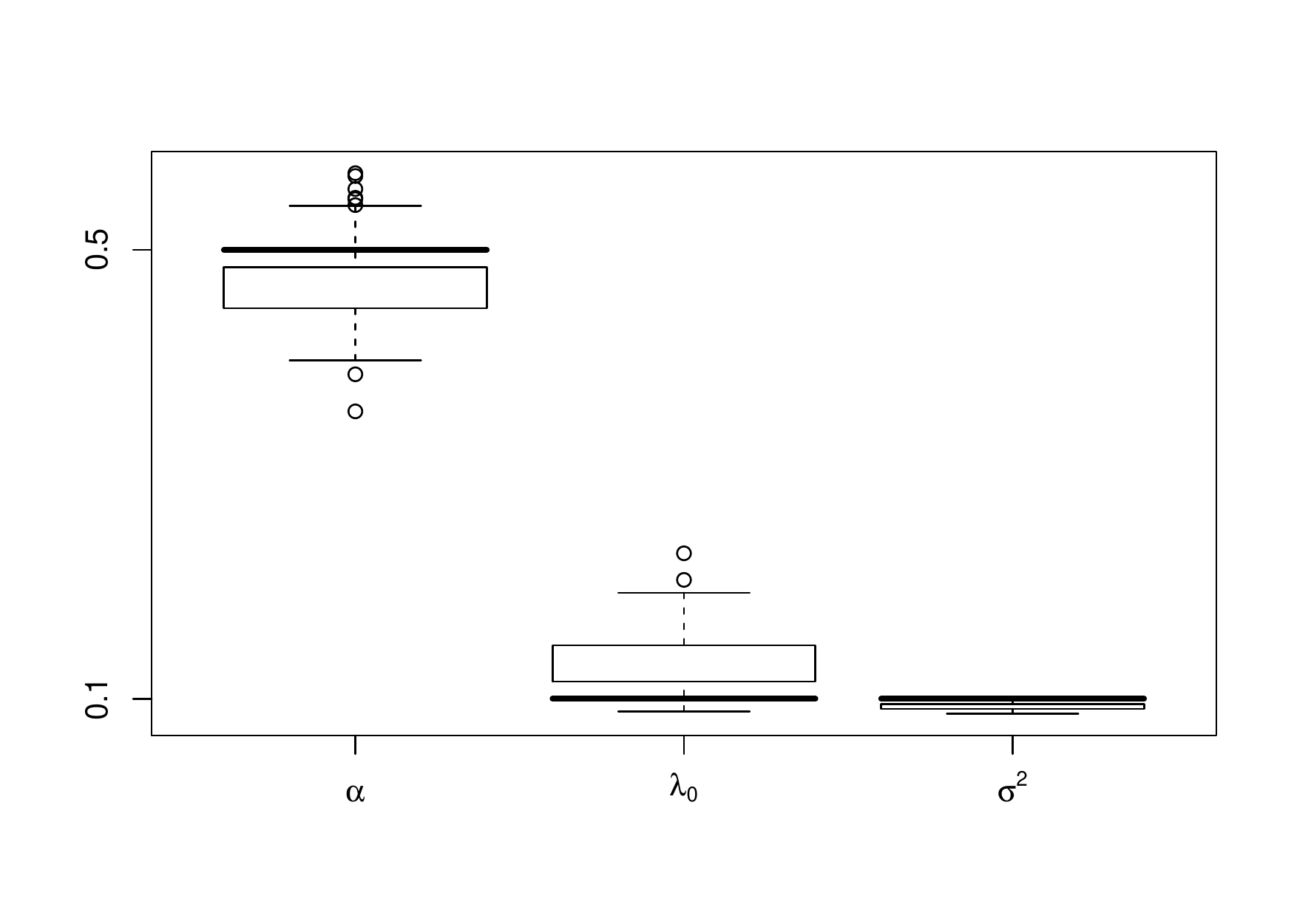}
	\caption{Parameter estimates for $\alpha, \lambda_0$ and $\sigma^2$ in Simulation 1 - thick lines indicate true values}
	\label{fig_s1_als}
\end{figure}

\subsubsection{Simulation 2: High dimensional setting}
\begin{figure*}[h]
	\centering
	\includegraphics[scale=0.8]{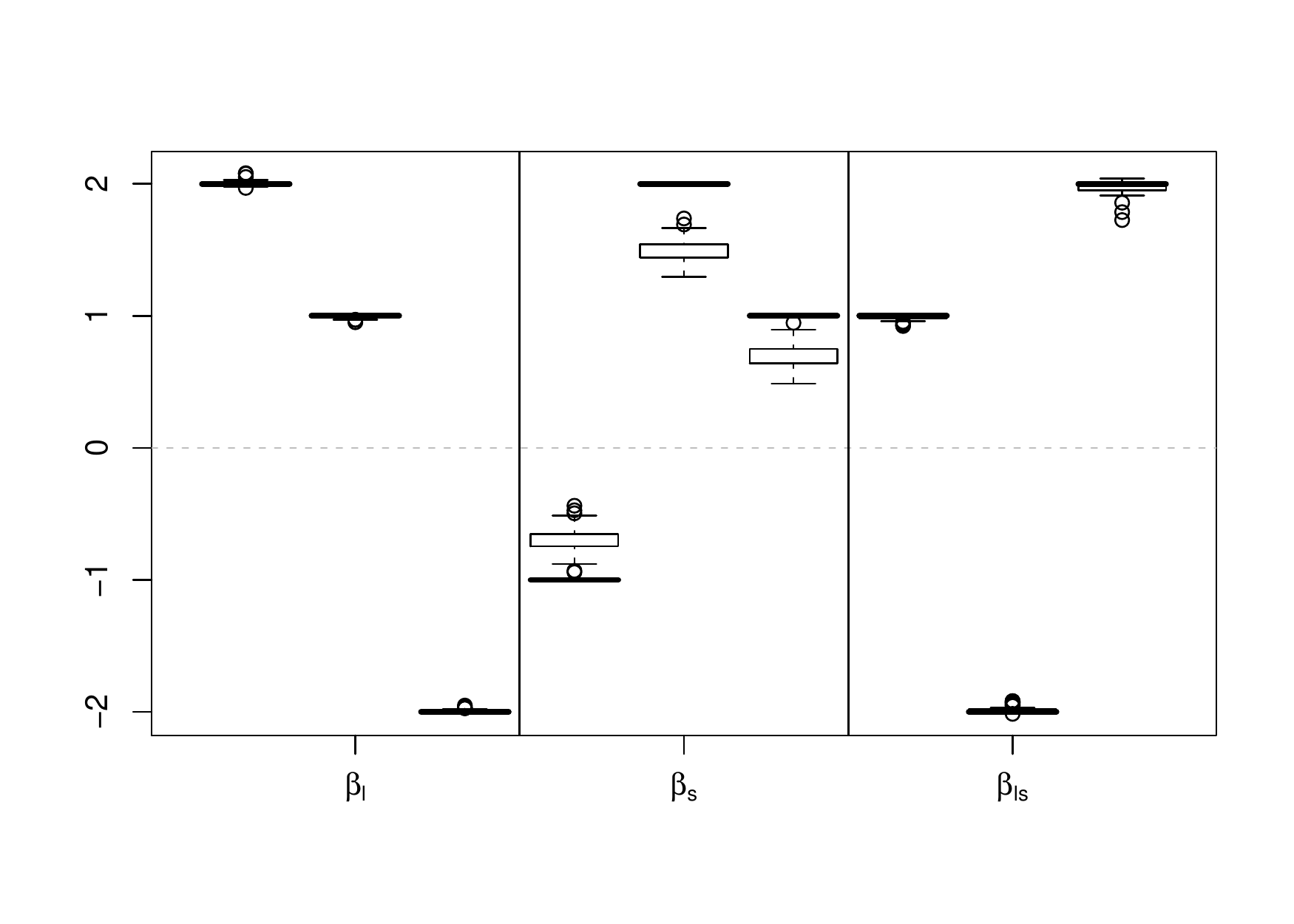}
	\caption{Coefficient estimates for 100 runs in Simulation 2 - thick lines indicate true values}
	\label{fig_s2_coef}
\end{figure*}

In the high dimensional setup we used the same parameters as S1 discussed in section \ref{sssec_low_dimensional} but now with 731 non-informative covariates per predictor overall leading to $p = 2202$. Technically, up to $n\cdot n_i = 2500$ longitudinal measurements are thinkable, but since the death of individuals leads to removal of the remaining values, the case $p<n\cdot n_i$ is highly unlikely and never occurred in the simulations, hence the number of covariates exceeded the number of measurements in every single simulation run. The coefficient estimates are visualized in Figure \ref{fig_s2_coef}. The coefficients in $\es$ experience higher shrinkage than in the low dimensional case, which also increases the trade-off between $\alpha$ and $\lambda_0$, see Table \ref{table_coef}. This is due to the fact, that the loss function of the survival part has a higher complexity than the quadratic loss appearing in the longitudinal part. The coefficients in the shared predictor on the other hand are not shrunken. The shared gradient $\boldsymbol{u}_\text{ls}$ is mainly driven by the quadratic loss, since the longitudinal measurements clearly outrun the survival data in numbers.
Variable selection performs way better in the high dimensional setup. While informative variables are picked $100\%$ of the time, false positives occur at rate $1.3\%$ in the longitudinal, $3.5\%$ in the survival and $2.2\%$ in the shared predictor. The averaged stopping iterations are
\begin{equation*}
	m_\text{l} = 148.5, \quad m_\text{s} = 157.75, \quad m_\text{ls} = 145.5.
\end{equation*}
The low value of $m_\text{s}$ compared to the low dimensional setup explains the relatively high shrinkage in the survival predictor.

\section{Example with AIDS data}\label{sec_data}
We test our boosting algorithm with the AIDS dataset consisting of 467 patients with advanced human immunodeficiency virus infection during antiretroviral \linebreak treatment in Abrams \textit{et al.} \cite{Abrams.1994}. Aim of the study was to compare two antiretroviral drugs indexed by \texttt{ddC} and \texttt{ddI} with respect to survival time. After study entry, patients had several follow-ups after 2, 6, 12 and 18 months where amongst other things the CD4 cell count was recorded. By the end of the study, 188 (40.26\%) of the patients had died. In addition we have the baseline covariates \texttt{gender}, \texttt{prev} (whether or not they had acquired immunodeficiency syndrome) and \texttt{azt} (whether some previous treatment failed or the patient was immune). More details regarding the study can be found in \cite{Abrams.1994}, the dataset itself is quiet popular and included in various \texttt{R} packages, e.g. \texttt{JM}. Since the CD4 cell count is a time-varying covariate, we formulate the following joint model
\begin{align*}
	y_{ij} &= \beta_0 + \beta_t t + \gamma_{0i} + \gamma_{1i} t + \beta_\text{ls}\cdot\texttt{ddI}_i + \varepsilon_{ij}\\
	&= m_i(t) + \varepsilon_{ij},\\
	\lambda_i(t) &= \lambda_0 \exp(\alpha m_i(t)\\
	&\quad  + \beta_{\text{s}1}\cdot\texttt{gender}_i + \beta_{\text{s}2}\cdot\texttt{azt}_i + \beta_{\text{s}3}\cdot\texttt{prev}_i)
\end{align*}
as we are interested in investigating the association between a time-dependent marker and the instantaneous risk for the death. To ensure the model contains random effects, we start with a random intercept estimated via maximum likelihood and $\gamma_{1i}^{[0]} = 0$, $i = 1,\dots,n$. The grid search was done with equally picked steplengths $\nu_\text{l} =  \nu_\text{s} = \nu_\text{ls} = 0.1$ on the grid defined by $m_\text{l}\in\{20,40,\dots,200\}$, $m_\text{s} \in \{30,60,\dots,300\}$ and $m_\text{ls} \in \{15,30,\dots,150\}$, since in the case $(200,300,150)$ convergence of the model would have been obtained. This leads to a total of 1000 estimated models where the best performing triple based on 10 fold cross validation is
\begin{equation*}
	m_\text{l} = 20, \quad m_\text{s} = 300, \quad m_\text{ls} = 120.
\end{equation*}
\begin{figure*}[h]
	\centering
	\includegraphics[scale=0.6]{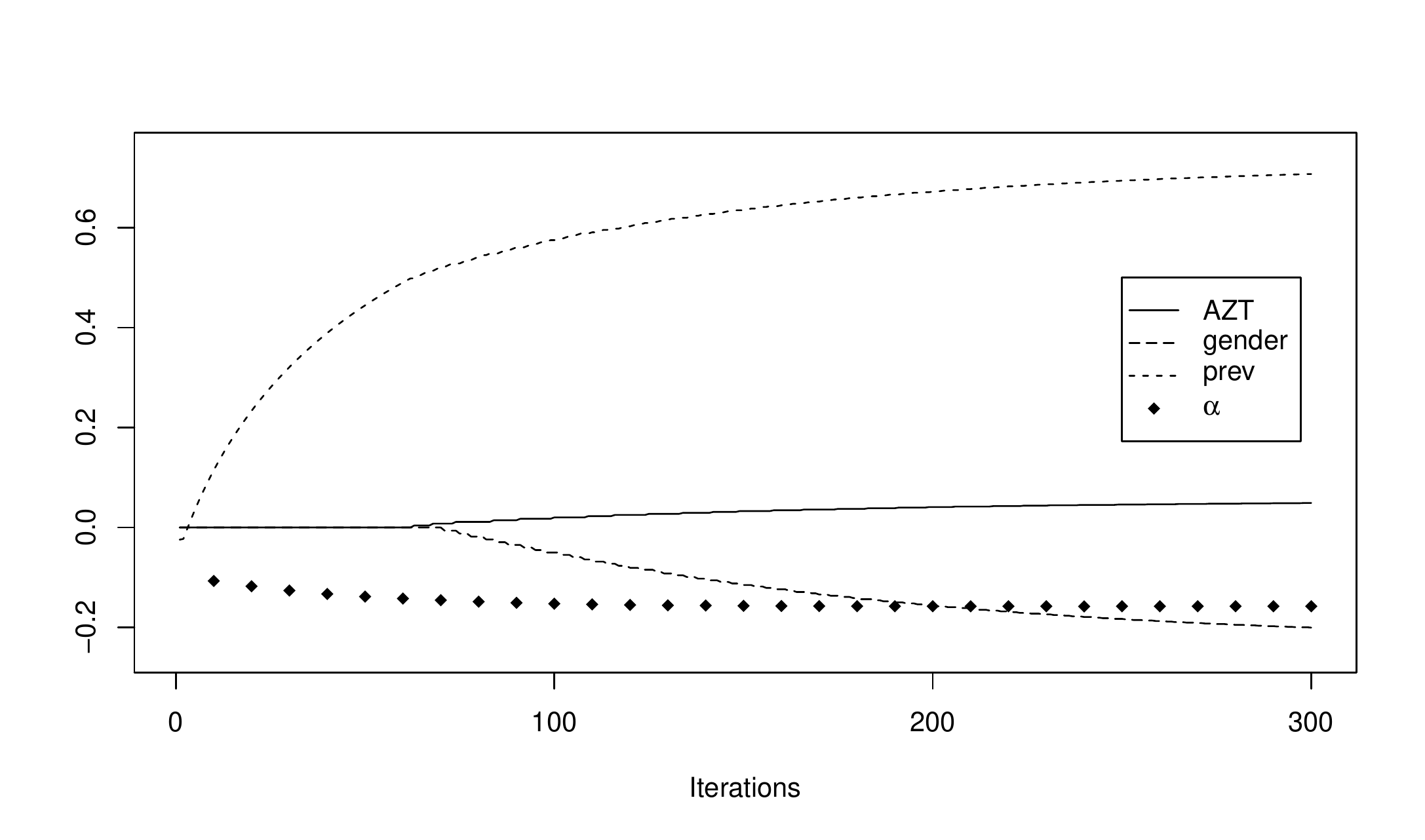}
	\caption{AIDS data example: coefficient progression in the survival submodel.}
	\label{fig_aids_path}
\end{figure*}
The coefficient paths of the survival model with this specific triple of stopping iterations are depicted in Figure \ref{fig_aids_path}. The algorithm picks the variable \texttt{prev} right away and therefore sees an unsurprisingly high linkage between suffering from AIDS and an increased risk for death. While \texttt{azt} is left fairly untouched by the algorithm, also \texttt{gender} gets picked after a while. The step-wise optimized association parameter takes the value $\alpha = -0.158$ indicating a negative influence of the CD4 cell count on the hazard ratio, thus a decrease of CD4 cells leads to a higher risk for death, which corresponds to the facts as to retrace in \cite{omim.cd4}.

\section{Discussion and Outlook}
In addition to the work by Waldmann \textit{et al.} \cite{Waldmann.2017} where longitudinal analysis was improved by incorporating information gained through the observed event-times, the extended algorithm offers a new approach to survival analysis by combining the advantages of both joint modeling and gradient boosting. Especially for high dimensional data this turns out to be a powerful tool, since gradient boosting not only makes joint modeling of high dimensional data manageable at all, but in addition also offers good variable selection properties. In the application the algorithm builds a stepwise coefficient progression allowing the algorithm to stop early in order to prevent overfitting and minimize the predictive risk.

Still, in its current form the simplicity of the model is not directly applicable to a more general class of datasets. In further work we plan to not only extend the baseline hazard $\lambda_0(t)$ to be time-varying, but also include a wide range of smooth effects in the longitudinal as well as in the survival submodel. Particularly in the latter this is quite a challenge, since gradient boosting is not capable of estimating time-varying effects in survival analysis. Therefore we intend to move on to likelihood-based boosting techniques \cite{Tutz.2006}, which address exactly these issues and already have proven to be practicable in very flexible survival models \cite{Hofner.2013}.

Another aspect of a possible extension would focus on including an allocation mechanism, where the algorithm allocates single covariates to each predictor instead of boosting three different predictor functions with predetermined covariates. This not only improves the model's flexibility but also reduces the computational effort by a lot, since the grid search is simplified to just one overall $m_\text{stop}$. A similar concept to boosting in multiple dimensions has been carried out by Thomas \textit{et al} \cite{Thomas.2017}. The allocation approach for joint models is being investigated in \cite{Waldmann.2018} and can also be extended to the present model.

Furthermore the variable selection properties with focus on the rate of false positives in low dimensional setups leaves room for improvement. Since the algorithm is not specifically designed for variable selection, some other possible solutions might be helpful. Stability selection \cite{Meinshausen.2010} being one of them already proved to be useful for boosting longitudinal and survival models, see \cite{Hofner.2015, Mayr.2016}. Probing on the other hand leads to better results regarding variable selection as investigated in \cite{Hepp.2017}. The idea of probing is to add non-informative phantom variables to each predictor and stop the boosting procedure, as soon as a phantom variable gets selected.

In this article the inference scheme proposed in \cite{Waldmann.2017} has been extended to a more general class of joint models, which allow way more focus on survival analysis. Nevertheless the extensions of manifold classes of joint models via tools from statistical learning are still far beyond its potential limits and open to further investigation.

\bibliographystyle{amsplain}
\bibliography{bibliography}

\providecommand{\bysame}{\leavevmode\hbox to3em{\hrulefill}\thinspace}
\providecommand{\MR}{\relax\ifhmode\unskip\space\fi MR }
\providecommand{\MRhref}[2]{%
  \href{http://www.ams.org/mathscinet-getitem?mr=#1}{#2}
}
\providecommand{\href}[2]{#2}
\begin{thebibliography}{10}

\bibitem{Abrams.1994}
Donald~I. Abrams, Anne~I. Goldman, Cynthia Launer, Joyce~A. Korvick, James~D.
  Neaton, Lawrence~R. Crane, Michael Grodesky, Steven Wakefield, Katherine
  Muth, Sandra Kornegay, David~L. Cohn, Allen Harris, Roberta Luskin-Hawk,
  Norman Markowitz, James~H. Sampson, Melanie Thompson, and Lawrence Deyton,
  \emph{A comparative trial of didanosine or zalcitabine after treatment with
  zidovudine in patients with human immunodeficiency virus infection.}, New
  England Journal of Medicine \textbf{330} (1994), 657--662.

\bibitem{Breiman.1998}
Leo Breiman, \emph{Arcing classifiers (with discussion)}, Ann. Statist.
  \textbf{26} (1998), 801--849.

\bibitem{Breiman.1999}
\bysame, \emph{Prediction games and arcing algorithms}, Neural Computation
  \textbf{11} (1999), 1493--1517.

\bibitem{Buehlmann.2007}
Peter B{\"u}hlmann and Torsten Hothorn, \emph{Boosting algorithms:
  Regularization, prediction and model fitting}, Statistical Science
  \textbf{22} (2007), no.~4, 477--505.

\bibitem{Buehlmann.2003}
Peter B{\"u}hlmann and Bin Yu, \emph{Boosting with the l2 loss}, Journal of the
  American Statistical Association \textbf{98} (2003), no.~462, 324--339.

\bibitem{Freund.1996}
Yoav Freund and Robert~E. Schapire, \emph{Experiments with a new boosting
  algorithm}, Proceedings of the Thirteenth International Conference on Machine
  Learning Theory, Morgan Kaufmann, San Francisco, 1996, pp.~148--156.

\bibitem{Guo.2004}
X.~Guo and B.~P. Carlin, \emph{Separate and joint modeling of longitudinal and
  event time data using standard computer packages}, The American Statistician
  \textbf{58} (2004), 16--24.

\bibitem{Hepp.2017}
Tobias Hepp, Janek Thomas, Andreas Mayr, and Bernd Bischl, \emph{Probing for
  sparse and fast variable selection with model-based boosting}, Computational
  and Mathematical Methods in Medicine \textbf{2017} (2017), 422--430.

\bibitem{Hofner.2008}
Benjamin Hofner, \emph{Variable selection and model choice in survival models
  with time-varying effects}, Diploma thesis, Ludwig-Maximilians-Universität
  München, 2008.

\bibitem{Hofner.2015}
Benjamin Hofner, Luigi Boccuto, and Markus Göker, \emph{Controlling false
  discoveries in high-dimensional situations: boosting with stability
  selection.}, BMC Bioinformatics \textbf{16} (2015), no.~1-2.

\bibitem{Hofner.2013}
Benjamin Hofner, Torsten Hothorn, and Thomas Kneib, \emph{Variable selection
  and model choice in structured survival models}, Computational Statistics
  \textbf{28} (2013), 1079--1101.

\bibitem{Hofner.2011}
Benjamin Hofner, Torsten Hothorn, Thomas Kneib, and Matthias Schmid, \emph{A
  framework for unbiased model selection based on boosting}, Journal of
  Computational and Graphical Statistics \textbf{20} (2011), 956--971.

\bibitem{Hofner.2014}
Benjamin Hofner, Andreas Mayr, Nikolay Robinzonov, and Matthias Schmid,
  \emph{Model-based boosting in r: A hands-on tutorial using the r package
  mboost}, Computational Statistics \textbf{29} (2014), no.~1-2, 3--35.

\bibitem{Mason.1999}
Llew Mason, Jonathan Baxter, Peter Bartlett, and Marcus Frean, \emph{Functional
  gradient techniques for combining hypotheses}, Advances in Large Margin
  Classifiers (1999).

\bibitem{Mayr.2012}
Andreas Mayr, Nora Fenske, Benjamin Hofner, Thomas Kneib, and Matthias Schmid,
  \emph{Generalized additive models for location, scale and shape for high
  dimensional data-a flexible approach based on boosting}, Journal of the Royal
  Statistical Society: Series C (Applied Statistics) \textbf{61} (2012), no.~3,
  403--427.

\bibitem{Mayr.2016}
Andreas Mayr, Benjamin Hofner, and Matthias Schmid, \emph{Boosting the
  discriminatory power of sparse survival models viaoptimization of the
  concordance index and stability selection.}, BMC Bioinformatics \textbf{17}
  (2012).

\bibitem{Meinshausen.2010}
Nicolai Meinshausen and Peter Bühlmann, \emph{Stability selection}, Journal of
  the Royal StatisticalSociety \textbf{72} (2010), 417--473.

\bibitem{omim.cd4}
{Online Mendelian Inheritance in Man, OMIM}, \emph{{CD4} antigen. {MIM} number:
  186940}, 2016.

\bibitem{Rizopoulos.2010}
Dimitris Rizopoulos, \emph{{JM}: An {R} package for the joint modelling of
  longitudinal and time-to-event data}, Journal of Statistical Software
  \textbf{35} (2010), no.~9.

\bibitem{Rizopoulos.2012}
\bysame, \emph{Joint models for longitudinal and time-to-event data: With
  applications in r}, Chapman {\&} Hall / CRC biostatistics series, vol.~6,
  {CRC Press}, Boca Raton, 2012.

\bibitem{Schmid.2008}
Matthias Schmid and Torsten Hothorn, \emph{Boosting additive models using
  component-wise p-splines}, Computational Statistics and Data Analyses
  \textbf{53} (2008), no.~2, 298--311.

\bibitem{Sweeting.2011}
Michael~J. Sweeting and Simon~G. Thompson, \emph{Joint modelling of
  longitudinal and time-to-event data with application to predicting abdominal
  aortic aneurysm growth and rupture}, Biometrical Journal \textbf{53} (2011),
  no.~5, 750--763.

\bibitem{Thomas.2017}
Janek Thomas, Andreas Mayr, Bernd Bischl, Matthias Schmid, Adam Smith, and
  Benjamin Hofner, \emph{Gradient boosting for distributional regression:
  faster tuning and improved variable selection via noncyclical updates},
  Statistics and Computing (2017), 1--15.

\bibitem{Tsiatis.2004}
Anastasios~A. Tsiatis and Marie Davidian, \emph{Joint modeling of longitudinal
  and time-to-event data: An overview}, Statistica Sinica \textbf{14} (2004),
  no.~3, 809--834.

\bibitem{Tutz.2006}
Gerhard Tutz and Harald Binder, \emph{Generalized additive modeling with
  implicit variable selection by likelihood-based boosting}, Biometrics
  \textbf{62} (2006), no.~4, 961--971.

\bibitem{Waldmann.2018}
Elisabeth Waldmann, Colin Griesbach, and Andreas Mayr, \emph{Variable selection
  and allocation in joint models for longitudinal and time-to-event data via
  boosting}, in preparation, 2018.

\bibitem{Waldmann.2017}
Elisabeth Waldmann, David Taylor-Robinson, Nadja Klein, Thomas Kneib, Tania
  Pressler, Matthias Schmid, and Andreas Mayr, \emph{Boosting joint models for
  longitudinal and time-to-event data}, Biometrical journal (2017).

\bibitem{Wolfe.1969}
Philip Wolfe, \emph{Convergence conditions for ascent methods}, SIAM Review
  \textbf{11} (1969), no.~2, 226--235.

\bibitem{Wulfsohn.1997}
Michael~S. Wulfsohn and Anastasios~A. Tsiatis, \emph{A joint model for survival
  and longitudinal data measured with error}, Biometrics \textbf{53} (1997),
  no.~1, 330.

\end{thebibliography}

\clearpage

\appendix
\section{Boosting algorithm}
	\begin{itemize}
		\item \textbf{Initialize} predictors $\hat{\eta}_\text{l}^{[0]}, \hat{\eta}_\text{s}^{[0]}$, $\hat{\eta}_\text{ls}^{[0]}$ and define base-learners $(h_{\text{l}1},\dots,h_{\text{l}p_\text{l}})$, $ (h_{\text{s}1},\dots,h_{\text{s}p_\text{s}})$ and $(h_{\text{ls}1},\dots,h_{\text{ls}p_\text{ls}})$, specify $h_\text{r}$ for the random effects and $h_t$ for the time effect. Initialize association parameter $\hat{\alpha}^{[0]}$, baseline hazard $\hat{\lambda}^{[0]}$ and model error $\hat{\sigma}^{2[0]}$. Choose stopping iterations $m_\text{stop,l}$, $m_\text{stop,s}$ and $m_\text{stop,ls}$ with overall maximum $m_\text{stop}$ and steplengths $\nu_\text{l}$, $\nu_\text{s}$ and $\nu_\text{ls}$.
		
		\item \textbf{for} $m=1$ to $m_\text{stop}$ \textbf{do}
		
		\item[] \textbf{step1: Update longitudinal predictor}\\
		\textbf{if} $m > m_\text{stop,l}$ set $\hat{\beta}_\text{l}^{[m]} = \hat{\beta}_\text{l}^{[m-1]}$, $\hat{\beta}_t^{[m]} = \hat{\beta}_t^{[m-1]}$, $\hat{h}_\text{r}^{[m]} = \hat{h}_\text{r}^{[m-1]}$ and skip this step\\
		\textbf{else}
		\begin{itemize}
			\item Compute $\boldsymbol{u}^{[m]}_{\text{l}}$ as
			\begin{align*}
			\boldsymbol{u}^{[m]}_{\text{l}} = \left(u^{[m]}_{\text{l}ij}\right)_{i=1,...,N, j=1,\dots,n_i}
			= \frac1{\hat{\sigma}^{2[m-1]}}\left(y_{ij} - \hat{\eta}^{[m-1]}_{\text{l}ij} - \hat{\eta}^{[m-1]}_{\text{ls}ij} \right)_{i=1,...,n, j=1,\dots,n_i}.
			\end{align*}
			\item Fit the negative gradient vector $\boldsymbol{u}^{[m]}_{\text{l}}$ separately to every base-learner specified for the longitudinal predictor $\el$:
			$$\boldsymbol{u}^{[m]}_{\text{l}} \xrightarrow{\text{base-learner}} \hat{h}_{\text{l}j}^{[m]}, \quad j\in\{1,\dots,p_\text{l},\text{r},t\}.$$
			\item Select the component
			$$j^* = \underset{j}\argmin \ \sum_{i=1}^n \sum_{j=1}^{n_i} (u_{\text{l}ij}^{[m]} - \hat{h}_{\text{l}j}^{[m]})^2$$
			that best fits $\boldsymbol{u}^{[m]}_{\text{l}}$ and update $\hat{\eta}_\text{l}^{[m]} = \hat{\eta}_\text{l}^{[m-1]} + \nu_\text{l}\hat{h}_{\text{l}j^*}^{[m]}$.
		\end{itemize}
		
		\item[] \textbf{step2: Update survival predictor}\\
		\textbf{if} $m > m_\text{stop,s}$ set $\hat{\beta}_\text{s}^{[m]} = \hat{\beta}_\text{s}^{[m-1]}$, $\hat{\lambda}_0^{[m]} = \hat{\lambda}_0^{[m-1]}$ and skip this step\\
		\textbf{else}
		\begin{itemize}
			\item Compute $\boldsymbol{u}^{[m]}_{\text{s}}$ as
			$$\boldsymbol{u}^{[m]}_{\text{s}}=\left( \delta_i - \hat{\lambda}_0^{[m-1]}\exp(\hat{\eta}_{\text{s}i}^{[m-1]}) \frac{\exp(\hat{\alpha}^{[m-1]}\hat{\eta}_{\text{ls}i}^{[m-1]}(\cdot,T_i)) - \exp(\hat{\alpha}^{[m-1]}\hat{\eta}_{\text{ls}-ti})}{\hat{\alpha}^{[m-1]}(\hat{\beta}_t^{[m-1]} + \hat{\gamma}_{1i}^{[m-1]})}\right)_{i=1,\dots,n}.$$
			\item Fit the negative gradient vector $\boldsymbol{u}^{[m]}_{\text{s}}$ separately to every base-learner specified for the survival predictor $\es$:
			$$\boldsymbol{u}^{[m]}_{\text{s}} \xrightarrow{\text{base-learner}} \hat{h}_{\text{s}j}^{[m]}, \quad j=1,\dots,p_\text{s}.$$
			\item Select the component
			$$j^* = \underset{j}\argmin \ \sum_{i=1}^n (u_{\text{s}i}^{[m]} - \hat{h}_{\text{s}j}^{[m]})^2$$
			that best fits $\boldsymbol{u}^{[m]}_{\text{s}}$ and update $\hat{\eta}_\text{s}^{[m]} = \hat{\eta}_\text{s}^{[m-1]} + \nu_\text{s}\hat{h}_{\text{s}j^*}^{[m]}$.
		\end{itemize}
		
		\item[] \textbf{step3: Update shared predictor}\\
		\textbf{if} $m > m_\text{stop,ls}$ set $\hat{\beta}_\text{ls}^{[m]} = \hat{\beta}_\text{ls}^{[m-1]}$ and skip this step\\
		\textbf{else}	
		\begin{itemize}
			\item Compute $\boldsymbol{u}^{[m]}_{\text{ls}}$ as
			\begin{align*}
			\boldsymbol{u}^{[m]}_{\text{ls}} &= \Bigg( \frac{y_{ij} - \hat{\eta}^{[m]}_{\text{l}ij} - \hat{\eta}^{[m-1]}_{\text{ls}i}}{\hat{\sigma}^{2[m-1]}} + \delta_i \hat{\alpha}^{[m-1]}\\
			&\quad\quad - \hat{\lambda}_0^{[m-1]}\exp(\hat{\eta}_\text{s}^{[m]}) \frac{\exp(\hat{\alpha}^{[m-1]}\hat{\eta}_{\text{ls}i}^{[m-1]}(\cdot,T_i)) - \exp(\hat{\alpha}^{[m-1]}\hat{\eta}_{\text{ls}-ti})}{\hat{\beta}_t^{[m-1]} + \hat{\gamma}_{1i}^{[m-1]}}\Bigg)_{i=1,...,n, j=1,\dots,n_i}.
			\end{align*}
			\item Fit the negative gradient vector $\boldsymbol{u}^{[m]}_{\text{ls}}$ separately to every base-learner specified for the shared predictor $\els$:
			$$\boldsymbol{u}^{[m]}_{\text{ls}} \xrightarrow{\text{base-learner}} \hat{h}_{\text{ls}j}^{[m]}, \quad j=1,\dots,p_\text{ls}.$$
			\item Select the component
			$$j^* = \underset{j}\argmin \ \sum_{i=1}^n \sum_{j=1}^{n_i} (u_{\text{ls}ij}^{[m]} - \hat{h}_{\text{ls}j}^{[m]})^2$$
			that best fits $\boldsymbol{u}^{[m]}_{\text{ls}}$ and update $\hat{\eta}_\text{ls}^{[m]} = \hat{\eta}_\text{ls}^{[m-1]} + \nu_\text{ls}\hat{h}_{\text{ls}j^*}^{[m]}$.
		\end{itemize}
		
		\item[] \textbf{step4: Update $\alpha$ and $\sigma^2$}
		\begin{itemize}
			\item \textbf{if} $m>m_\text{stop,l}\wedge m>m_\text{stop,ls}$ set $\hat{\sigma}^{2[m]}=\hat{\sigma}^{2[m-1]}$ and skip this update\\
			\textbf{else} for the number of total longitudinal observations $N = \sum_i n_i$ and covariates $p$ set
			$$\sigma^{2[m]} = \frac{1}{N - p} \sum_{i=1}^n \sum_{j=1}^{n_i} (y_{ij} - \hat{\eta}_{\text{l}ij}^{[m]} - \hat{\eta}_{\text{ls}ij}^{[m]})^2$$.
			\item \textbf{if} $m>m_\text{stop,ls}$ set $\hat{\alpha}^{[m]}=\hat{\alpha}^{[m-1]}$ and skip this update\\
			\textbf{else}
			$$ \hat{\alpha}^{[m]} = \underset{\alpha\in\R}\argmax \ \ell(\hat{\eta}_\text{l}^{[m]},\hat{\eta}_\text{s}^{[m]},\hat{\eta}_\text{ls}^{[m]},\alpha,\hat{\lambda}_0^{[m]},\hat{\sigma}^{2[m]}| \by, \bT, \bd)$$	
		\end{itemize}
		
		\item \textbf{end} when $m = m_\text{stop}$.
	\end{itemize}	

\section{Simulation algorithm}\label{seq_algo_sim}

	\begin{itemize}
		\item \textbf{Choose} values for $n, n_i, \beta_0, \bl, \bs, \bls, \beta_t, \alpha, \lambda_0$ and $\sigma^2$.
		
		\item \textbf{Generate} $n\cdot n_i$ time points $\bt$ for longitudinal measurements the following way:
		\begin{itemize}
			\item Sample $d_{ij} \sim \mathcal{U}(\{1,\dots,365\})$ and set $\tilde{t}_{ij} := (j-1)\cdot 365 + d_{ij}$ for $i = 1,\dots,n$ and $j=1,\dots,n_i$.
			\item For each $i$ shift observation times to the left, so we have $t_{i1} = 0$. 
			\item Standardize time points to the unit interval by $t_{ij} := \tilde{t}_{ij}/(n_i \cdot 365)$
			\item Set $\bt := (t_{ij})_{i=1,\dots,n,j=1,\dots,n_i}$ as the collection of all longitudinal time points.
		\end{itemize}
		
		\item \textbf{Generate} covariate matrices $\Xl, \Xs, \Xls$ using uniformly distributed random numbers and random effects $(\gamma_{0i},\gamma_{1i})_{i=1,\dots,n}$ following a normal distribution with mean 0.
		
		\item \textbf{Calculate} the predictor vectors
		\begin{equation*}
		\bel = \beta_0 + \Xl\bl, \quad \bes = \Xs\bs, \quad \bels = \Xls\bls + \gamma_0 + (\beta_t + \gamma_1) \bt		
		\end{equation*}
		
		\item \textbf{Simulate} longitudinal measurements $\by = (y_{ij})$ by
		\begin{equation*}
		\by := \mathcal{N}^{\otimes n\cdot n_i}(\bel + \bels, \sigma^2\cdot I_{n\cdot n_i}) 
		\end{equation*}
		where $\mathcal{N}^{\otimes n\cdot n_i}$ denotes the $n\cdot n_i$-dimensional multivariate normal distribution and $I_{n\cdot n_i}$ the corresponding unit matrix.
		
		\item \textbf{Draw} event times by generating random numbers $u_i \sim \mathcal{U}([0,1])$ and setting
		\begin{equation*}
		T_i^* := \frac{\log\left(\frac{-\log(1-u_i)\alpha(\beta_t + \gamma_{1i})}{\lambda_0 \exp(\eta_{\text{s}i})} + \exp(\alpha\eta_{\text{ls}-ti})\right) - \alpha(\eta_{\text{ls}-ti})}{\alpha (\beta_t + \gamma_{1i})}
		\end{equation*}
		according to inversion sampling. Set $T_i := \max(T_i^*, t_{in_i})$ to obtain censored data with censoring indicator $\delta_i := I(T_i^* \leq t_{in_i})$ and receive the \textit{observed} survival outcome $(\bT,\bd)=(T_i,\delta_i)_{i=1,\dots,n}$.
		
		\item \textbf{Delete} all entries from $\by, \Xl, \Xs, \Xls$ corresponding to times $t_{ij}>T_i$ for every individual $i$.
		
		\item \textbf{Add} columns with i.i.d. generated random numbers as non-informative covariates to $\Xl, \Xs$ and $\Xls$.
	\end{itemize}

\section{Gradient of the survival predictor}

Since computing the functional derivative with respect to the longitudinal predictor is straightforward and the computation of the shared gradient analogous to the explanation in the appendix of \cite{Waldmann.2017}, we will sketch out the calculus of the functional derivative with respect to the survival predictor.

As mentioned in section \ref{sec_boosting}, we set the negative log-likelihood as our loss function, thus $\rho=-\ell$. Since $\es$ appears only in the survival component of the likelihood we get
\begin{align*}
\boldsymbol{u}_\text{s} &= - \frac{\partial\rho}{\partial\es} = \frac{\partial\ell}{\partial\es}\\
&= \frac{\partial}{\partial\es}\left( \delta\log\lambda_0 + \delta\alpha\els + \delta\es - \lambda_0\int_0^T \exp\left(\es + \alpha\els\right) dt \right)\\
&= \delta - \lambda_0\int_0^T \exp\left(\es + \alpha\els\right) dt\\
&= \delta-\lambda_0\exp(\es)\int_0^T\exp(\alpha\els)dt,
\end{align*}
as $\es$ does not depend on $t$. We get
\begin{equation*}
\int_0^T\exp(\alpha\els)dt = \frac{\exp(\alpha\els(\cdot,T)) - \exp(\alpha\eta_{\text{ls}-t})}{\alpha(\beta_t + \gamma_1)}
\end{equation*}
with $\eta_{\text{ls}-t}$ again denoting the time-constant part of $\els$, so we have
\begin{equation*}
\boldsymbol{u}_\text{s} = \delta - \lambda_0\exp(\es) \frac{\exp(\alpha\els(\cdot,T)) - \exp(\alpha\eta_{\text{ls}-t})}{\alpha(\beta_t + \gamma_1)}
\end{equation*}
for the final gradient.

\end{document}